\documentclass[prb,preprint,showpacs,preprintnumbers,amsmath,amssymb,floatfix]{revtex4}

\usepackage[dvipdfmx]{graphicx}
\usepackage{bm}

\newcommand{\be}{\begin{equation}}
\newcommand{\ee}{\end{equation}}
\newcommand{\eq}[1]{Eq.~(\ref{#1})}

\def\bea{\begin{eqnarray}}
\def\eea{\end{eqnarray}}

\def\vq{{\bf q}}

\def\vk{{\bf k}}

\begin{document}

\title{Superconductivity with and without glue and the role of the 
double-occupancy forbidding constraint in the $t$-$J$-$V$ model}

\author{Luciano Zinni, Mat\'{\i}as Bejas, and Andr\'es Greco}

\affiliation{
Facultad de Ciencias Exactas, Ingenier\'{\i}a y Agrimensura and
Instituto de F\'{\i}sica Rosario
(UNR-CONICET).
Avenida Pellegrini 250-2000 Rosario-Argentina.
}

\date{\today}

\begin{abstract}
The occurrence of retarded (with glue) and unretarded (without glue) pairing is thoroughly
discussed in cuprates. We analyze some aspects of this problem in the context of the $t$-$J$-$V$ model in a 
large-$N$ approximation. When $1/N$ renormalizations are neglected the mean-field result is 
recovered, where the unretarded $d$-wave superconducting pairing triggered by the spin-exchange interaction $J$
is obtained. 
However, the presence 
of a non-negligible nearest-neighbors Coulomb interaction $V({\bf q})$ kills superconductivity.
If the non-double-occupancy constraint and its fluctuations are considered, the situation changes drastically. 
In this case, $V({\bf q})$ is screened making $d$-wave superconductivity very robust. 
In addition, we show that the early proposal for the presence of an unretarded pairing contribution
triggered by the spin-exchange interaction $J$ can be discussed in this context.
\end{abstract} 

\maketitle

\section{Introduction}

The origin of superconductivity in high-$T_c$ cuprates is under an intense 
debate since its discovery in 1986. Not only the high value of the 
superconducting critical temperature $T_c$ is surprising, but these materials are also anisotropic
and the metallic and superconducting properties show a two-dimensional character.
The phase diagram 
in the temperature  and doping  plane 
of these materials shows unconventional characteristics, 
as the dome-shaped behavior of $T_c$ against doping in the proximity 
to the antiferromagnetic insulator and the pseudogap 
phase at low doping (see Ref. [\onlinecite{keimer15}] for a review). In addition to these  
features, the superconducting gap has $d$-wave symmetry\cite{keimer15,timusk99}. 
All members of the cuprate family share similar characteristics, suggesting 
the existence of a universal physics. 

Phenomenological theories where pairing is due to antiferromagnetic fluctuations\cite{monthoux91,monthoux93}
were proposed for explaining the $d$-wave symmetry of the superconducting gap in 
the proximity to antiferromagnetism. In this scenario, the two-dimensional antiferromagnetic 
fluctuations play the role of a retarded glue, as 
phonons in conventional low-temperature superconductors. 

In the early times, the Hubbard and the $t$-$J$ models were recognized\cite{anderson87} as minimal 
microscopic models for cuprates. The Hubbard model treated in the framework of a 
weak coupling random phase approximation
shows $d$-wave superconductivity\cite{bulut96a}, where the effective pairing 
interaction is mediated by the dynamical spin susceptibility which acts as a pairing glue. 
In this approach, the nearest-neighbors Coulomb interaction,
$V(\vq)=2V[\cos(q_x)+\cos(q_y)]$, 
which is expected to be non-negligible
in cuprates\cite{feiner96}, affects superconductivity because it has 
a $d$-wave repulsive projection. It is therefore important to understand 
why $T_c$ remains large even if a nearest-neighbors Coulomb interaction is present. 
The same effect of $V(\vq)$ is expected  in the antiferromagnetic phenomenological 
theories\cite{monthoux91,monthoux93}. 

The study of $d$-wave superconductivity and the role of the 
nearest-neighbors Coulomb interaction in the Hubbard model is 
huge\cite{senechal13,esirgen99,husemann12,jiang18}.
In Ref. [\onlinecite{senechal13}] it was shown that $d$-wave superconductivity in the Hubbard model 
is almost unaffected by $V(\vq)$ if the strong coupling limit is properly treated.
In addition, since the two-dimensional Hubbard  model reduces 
to the $t$-$J$ model in the large-$U$ limit \cite{chao78}, the question about the role of $V(\vq)$
on superconductivity is also of interest in the $t$-$J$ model. On the other hand, 
retarded (with glue) and unretarded (without glue) 
paring\cite{anderson07,maier08} is under discussion in the $t$-$J$ model.
While in Ref. [\onlinecite{maier08}] pairing was discussed as composed by a retarded and an unretarded 
contributions, in Ref. [\onlinecite{anderson07}] only an unretarded pairing was considered as the relevant one.

At the mean-field level, the $t$-$J$ model shows $d$-wave 
superconductivity\cite{baskaran93} arising from the unretarded exchange interaction 
$J [\cos(q_x)+\cos(q_y)]$, where $J$ is the spin-exchange 
coupling. Since the exchange interaction has the same 
form of the nearest-neighbors Coulomb interaction, the last one is, in principle,  
detrimental to superconductivity. 
Using a large-$N$ approach on the $t$-$J$-$V$  model, in this paper we discuss superconductivity and the 
role of the nearest-neighbors Coulomb interaction $V(\vq)$. When the local constraint 
that prohibits double occupancy is not included 
superconductivity is strongly affected 
by $V(\vq)$, even for $V$ of the order of $J$. 
Including the constraint $d$-wave 
superconductivity is robust against $V(\vq)$, even for $V \gg J$. We also found that the leading contribution to 
superconductivity is mainly provided by the unretarded exchange, however, this 
contribution is efficient 
only if the constraint is properly
included. 
In Sec. II we present a summary of the formalism, in Sec. III the results and discussions,
and in Sec. IV the conclusions.

\section{Model and summary of the formalism}

As a minimal model, 
we study the $t$-$J$-$V$ model on a square lattice, 
\begin{equation}
H = -\sum_{\langle i,j \rangle,\sigma} t_{i j}\tilde{c}^\dag_{i\sigma}\tilde{c}_{j\sigma} + 
\sum_{\langle i,j \rangle} J_{ij} \left( \vec{S}_i \cdot \vec{S}_j - \frac{1}{4} n_i n_j \right)
+\sum_{{\langle i,j \rangle}} V_{ij} n_i n_j \, ,
\label{tJV}  
\end{equation}
where $\tilde{c}^\dag_{i\sigma}$ ($\tilde{c}_{i\sigma}$) is 
the creation (annihilation) operator of electrons with spin $\sigma (=\uparrow, \downarrow)$  
in the Fock space without double occupancy at any site,  
$n_i=\sum_{\sigma} \tilde{c}^\dag_{i\sigma}\tilde{c}_{i\sigma}$ 
is the electron density operator, $\vec{S}_i$ is the spin operator. 
The hopping (spin exchange) $t_{i j}$ ($J_{i j}$) takes the value $t$ ($J$) between the first nearest-neighbors
sites. 
$V_{ij}$ is a nearest-neighbors Coulomb interaction with strength $V$. 

It is non-trivial to study the $t$-$J$ model because of the local constraint 
that prohibits the double occupancy at any site. In addition, the operators involved in the 
$t$-$J$ model are Hubbard operators\cite{hubbard63} which satisfy non-standard commutation rules.
We employ here a large-$N$ technique based on a path integral representation in terms of the Hubbard 
operators (see Refs.~[\onlinecite{foussats02,foussats04}] and references therein). 
In the large-$N$ scheme, the number of spin components is extended from 2 to $N$ and the
physical quantities are computed in powers of $1/N$. In what follows the spin index $\sigma$ is called 
$p$.

In the framework of the large-$N$ path integral approach,
the $t$-$J$ model is mapped to an effective theory  
described in terms of fermions, bosons, and their mutual interactions\cite{foussats04}. 

\begin{figure}[htb]
\begin{center}
\setlength{\unitlength}{1cm}
\includegraphics[width=8cm,angle=0]{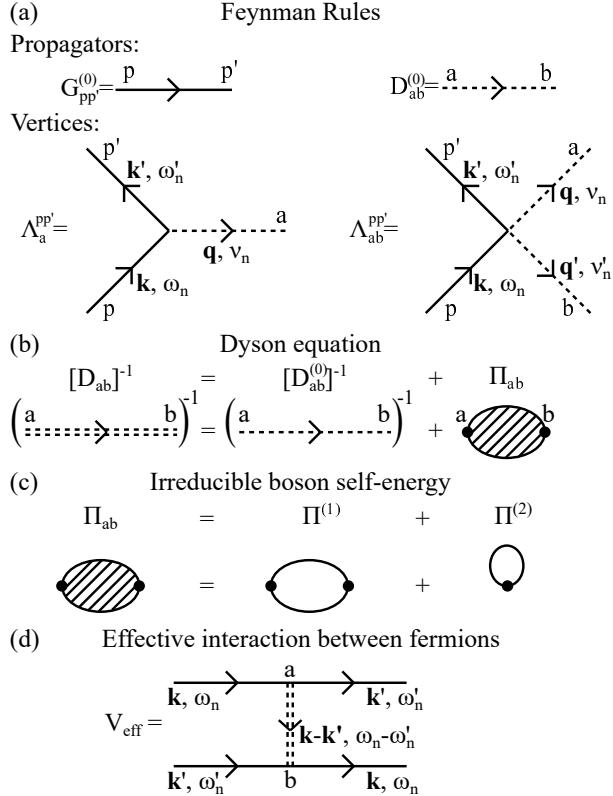}
\end{center}
\caption{(a) Summary of the Feynman rules. Solid line represents the fermionic propagator $G_{pp'}^{(0)}$.
Dashed line
represents the $6 \times 6$  bosonic propagator
$D_{ab}^{(0)}$. $\Lambda^{pp'}_a$ and $\Lambda^{pp'}_{ab}$ represent
the interaction between two fermions and one and two bosons, respectively. 
(b) Diagrammatic representation of the Dyson equation.
(c) The two different contributions to the irreducible boson self-energy.
(d) Effective interaction between fermions. Looking at the order of the propagators and vertices
we see that $V_{{\rm eff}}$ is $O(1/N)$, thus superconductivity arises 
at $O(1/N)$ in this large-$N$ scheme.
}
\label{fig:feynman}
\end{figure}

{\it a) Fermions:} 
We obtain a fermionic
propagator  [solid line in Fig. \ref{fig:feynman}(a)], 
\begin{equation}
G^{(0)}_{pp'} ({\bf k}, {\rm i}\omega_{n}) = \frac{\delta_{pp'}}{{\rm i}\omega_{n} -\varepsilon_{\vk}}\,,
\label{G0-A}
\end{equation}

\noindent with the electronic dispersion 
\be
\varepsilon_{\vk}= -2 \left( t \frac{\delta}{2}+\Delta \right) \left[\cos \left(k_{x}\right)+\cos \left(k_{y}\right)\right]- \mu \,.\\
\label{Epara-A}
\ee

For a given doping $\delta$, the chemical potential $\mu$ and $\Delta$ are determined self-consistently by solving
\be
1-\delta = \frac{2}{N_s} \sum_{\vk} n_F\left(\varepsilon_\vk\right) \; ,
\ee
\noindent and 
\bea{\label {Delta-A}} 
\Delta = \frac{J}{4N_s} \sum_{\vk}  \left[\cos \left(k_x\right) +\cos \left(k_y\right)\right] n_F\left(\varepsilon_\vk\right) \; , 
\eea

\noindent where $n_F$ is the Fermi function and $N_s$ is the total number of lattice sites. 
The momentum $\vk$  is measured in units of the inverse of the lattice constant. In \eq{G0-A} 
$\omega_{n}$ is a fermionic Matsubara frequency. The Green's function 
$G^{(0)}_{pp'}({\bf k}, {\rm i}\omega_{n})$ is $O(1)$ in the context of the $1/N$ expansion.

In the present formalism, the spin-exchange term or $J$-term of the $t$-$J$-$V$ model [\eq{tJV}] is treated by 
introducing a bond-field variable that describes charge fluctuations on the bond connecting nearest-neighbors sites 
along the $x$- and $y$-directions. $\Delta$ is the static mean-field value of this bond field. 
Although the electronic dispersion [\eq{Epara-A}] looks like that in a free electron system, 
the hopping integral $t$ is renormalized by doping $\delta$ 
because of electron-correlation effects. In addition, there is a contribution $\Delta$ which depends on $J$.

{\it b) Bosons:} 
We define a six-component bosonic field 
\begin{equation}
\delta X^{a} = (\delta
R\;,\;\delta{\lambda},\; r^{x},\;r^{y}
,\; A^{x},\;
A^{y})\, ,
\label{boson-field}
\end{equation}
where $\delta R$ describes the fluctuations of the number of holes at a given site, thus it 
is related to on-site charge fluctuations, $\delta \lambda$ is the fluctuation of the
Lagrange multiplier introduced to enforce the constraint that prohibits the double occupancy 
at a given site, and $r^{x}$ and $r^{y}$ ($A^{x}$ and $A^{y}$) 
describe fluctuations of the real (imaginary) part of the bond field coming 
from the $J$-term. 

The $6\times6$ bare bosonic propagator associated with $\delta X^{a}$ 
[dashed line in Fig. \ref{fig:feynman}(a)],
connecting two generic components $a$ and $b$ is
\begin{eqnarray}
\left[D^{(0)}_{ab}({\bf q},\mathrm{i}\nu_{n})\right]^{-1}= N \left(
 \begin{array}{cccccc}
\frac{\delta^2}{2} \left( V-\frac{J}{2}\right)\left[\cos \left(q_x\right)+\cos \left(q_y\right) \right]
& \frac{\delta}{2} & 0 & 0 & 0 & 0 \\
   \frac{\delta}{2} & 0 & 0 & 0 & 0 & 0 \\
   0 & 0 & \frac{4}{J}\Delta^{2} & 0 & 0 & 0 \\
   0 & 0 & 0 & \frac{4}{J}\Delta^{2} & 0 & 0 \\
   0 & 0 & 0 & 0 & \frac{4}{J}\Delta^{2} & 0 \\
   0 & 0 & 0 & 0 & 0 & \frac{4}{J}\Delta^{2} 
 \end{array}
\right)  \; ,
\label{D0-A}
\end{eqnarray}
\noindent where ${\bf q}$ and $\nu_n$ are the momentum and  bosonic Matsubara frequencies, respectively.
The factor $N$ shows that the $6 \times 6$ bosonic propagator $D^{(0)}_{ab}$ is $O(1/N)$, 
and it is frequency independent.
The element $(1,1)$ in \eq{D0-A} carries the information of $\frac{1}{4} J_{ij}  n_i n_j$ and 
$V_{ij} n_i n_j$ of \eq{tJV}, while the information of  $J_{ij} \vec{S}_i \cdot \vec{S}_j$
is contained in the elements $(a,a)$ with $a=3$-$6$.

{\it c) Interaction vertices:} 
For computing quantities up to $O(1/N)$ the present large-$N$ scheme leads to three-legs and 
four-legs vertices [Fig. \ref{fig:feynman}(a)]. 

The three-legs vertex 
\begin{eqnarray}\label{gamm3}
\Lambda^{pp'}_{a} =&&  (-1) \left[\frac{{\rm i}}{2}(\omega_n +
{\omega'}_n) + \mu + 2\Delta \sum_{\eta}
\cos \left( k_\eta-\frac{q_\eta}{2} \right) \cos\frac{q_\eta}{2};\;1\; ; 
- 2\Delta \cos \left( k_x-\frac{q_x}{2} \right) \; ; \right. \nonumber \\
&&  \left. 
- 2\Delta \cos \left(k_y-\frac{q_y}{2} \right) ; \;  
 2 \Delta \sin \left(k_x-\frac{q_x}{2} \right) ;\; 
 2 \Delta \sin \left(k_y-\frac{q_y}{2} \right) \right]  \delta^{pp'} \, ,
 \label{eq:vertex-lambda}
\end{eqnarray}
\noindent where $\eta = x$, $y$, represents the interaction between two fermions and one boson. 

The four-legs vertex $\Lambda^{pp'}_{ab}$  represents the interaction between
two fermions and two bosons.  
$\Lambda^{pp'}_{ab}$ fulfills the symmetry of 
$\Lambda^{pp'}_{ab} = \Lambda^{pp'}_{ba}$, and  
the only elements different from zero are:
\begin{eqnarray}
\Lambda^{pp'}_{\delta R \delta R}=&&  \left[\frac{{\rm i}}{2} (\omega_n + {\omega'}_n)
+ \mu  \right. \nonumber \\
&&  \left. + \Delta \sum_{\eta}
\cos \left( k_\eta-\frac{q_\eta+q'_\eta}{2} \right)\;
\left( \cos\frac{q_\eta}{2} \; \cos\frac{q'_\eta}{2}\;
+\; \cos\frac{q_\eta+q'_\eta}{2} \right) \right] \delta^{pp'}\, ,
\end{eqnarray}
\begin{equation}
\Lambda^{pp'}_{\delta R \delta \lambda}=\frac{1}{2}\delta^{pp'} \, ,
\end{equation}
\begin{equation}
\Lambda^{pp'}_{\delta R \, r^\eta}=-\Delta \cos \left( k_\eta-\frac{q_\eta+q'_\eta}{2} \right)\cos \frac{q'_\eta}{2}\, \delta^{pp'} \, ,
\end{equation}
and
\begin{equation}
\Lambda^{pp'}_{\delta R \, A^\eta}=\Delta \sin \left( k_\eta-\frac{q_\eta+q'_\eta}{2} \right)\cos \frac{q'_\eta}{2} \, \delta^{pp'} \, .
\end{equation}

\noindent Each vertex conserves momentum and energy and it is $O(1)$. For readability reasons we drop the frequencies and momenta in the left hand side of the definitions of the three- and four-legs vertices
$\Lambda^{pp'}_{a}$ and $\Lambda^{pp'}_{ab}$ [see Fig. \ref{fig:feynman}(a) 
for the frequency and momentum dependence].

By using the propagators and vertices summarized in Fig. \ref{fig:feynman}(a) we can draw Feynman diagrams as usual.

From the Dyson equation [Fig. \ref{fig:feynman}(b)],  
the bosonic bare propagator $D^{(0)}_{ab}$ is renormalized at $1/N$ order,
\be
[D_{ab}(\vq,\mathrm{i}\nu_n)]^{-1} 
= [D^{(0)}_{ab}(\vq,\mathrm{i}\nu_n)]^{-1} - \Pi_{ab}(\vq,\mathrm{i}\nu_n)\,,
\label{dyson-A}
\ee
where the $6 \times 6$ boson self-energy
matrix $\Pi_{ab}$ [Fig. \ref{fig:feynman}(c)] is: 
\begin{eqnarray}
&& \Pi_{ab}(\vq,\mathrm{i}\nu_n)
            = -\frac{N}{N_s}\sum_{\vk} h_a(\vk,\vq,\varepsilon_\vk-\varepsilon_{\vk-\vq}) 
            \frac{n_F(\varepsilon_{\vk-\vq})-n_F(\varepsilon_\vk)}
                                  {\mathrm{i}\nu_n-\varepsilon_\vk+\varepsilon_{\vk-\vq}} 
            h_b(\vk,\vq,\varepsilon_\vk-\varepsilon_{\vk-\vq}) \nonumber \\
&& \hspace{25mm} - \delta_{a\,1} \delta_{b\,1} \frac{N}{N_s}
                                       \sum_\vk \frac{\varepsilon_\vk-\varepsilon_{\vk-\vq}}{2}n_F(\varepsilon_\vk) \; ,
                                       \label{Pi-A}
\end{eqnarray}
\noindent with $h_a$ given by 
\begin{align}
 h_a(\vk,\vq,\nu) =& \left\{
                   \frac{2\varepsilon_{\vk-\vq}+\nu+2\mu}{2}+
                   2\Delta \left[ \cos\left(k_x-\frac{q_x}{2}\right)\cos\left(\frac{q_x}{2}\right) +
                                  \cos\left(k_y-\frac{q_y}{2}\right)\cos\left(\frac{q_y}{2}\right) \right];1;
                 \right. \nonumber \\
               & \left. -2\Delta \cos\left(k_x-\frac{q_x}{2}\right)\hspace{-0.1cm}; -2\Delta \cos\left(k_y-\frac{q_y}{2}\right)\hspace{-0.1cm};
                         2\Delta \sin\left(k_x-\frac{q_x}{2}\right)\hspace{-0.1cm};  2\Delta \sin\left(k_y-\frac{q_y}{2}\right)
                 \right\} \, .
\label{vertex-h-A}
\end{align}

The vertices $\Lambda^{pp'}_{a}$ and $\Lambda^{pp'}_{ab}$ not only represent interactions
from the Hamiltonian \eq{tJV} but, as they come from the path integral, they contain also contributions
from the algebra of the Hubbard operators and the non-double-occupancy constraint, which introduce
a frequency dependence.
Due to this frequency dependence, the computation of the first and second diagrams in Fig. \ref{fig:feynman}(c) leads to finite and infinite contributions.
However, the ghost fields from the Jacobian in the path integral give rise to terms that cancel exactly these
infinities\cite{foussats02,foussats04}.

The $6 \times 6$ dressed bosonic propagator $D_{ab}$ contains all possible charge fluctuations 
of the $t$-$J$ model on the square lattice, and all are treated on equal footing\cite{bejas12}.
The large-$N$ approach weakens the effective
spin interaction compared with the one associated with the charge degrees of freedom. $D_{ab}$ with $a,b=1,2$ describes 
on-site charge fluctuations associated to $\delta R$ and $\delta{\lambda}$.  
The presence of $\delta{\lambda}$ indicates that the non-double-occupancy constraint and its fluctuations are 
taken into account in the calculation.
The element $(1,1)$ of $D_{ab}$ is related to 
the usual charge-charge correlation function \cite{foussats02}. $D_{22}$ and $D_{12}$ correspond to fluctuations associated with the non-double-occupancy condition 
and correlations between non-double-occupancy condition and charge-density fluctuations, respectively.
We call this case as on-site charge 
sector or the $2\times 2$ sector. If $a,b=3$-$6$, $D_{ab}$ describes bond-charge fluctuations associated to 
$r^{x}$, $r^{y}$, $A^{x}$, and $A^{y}$. We call this case as the bond-charge sector or 
the $4\times 4$ sector. $D_{ab}$ also contains the mixing of both sectors, however it was shown 
that the coupling between on-site and bond-charge fluctuations is negligible\cite{bejas17}. 
If $J=0$ the $6 \times 6$ $D_{ab}$ reduces to the $2\times2$ sector, and only on-site charge fluctuations are involved. 

The superconducting effective interaction between fermions,
$V_{{\rm eff}}({\bf k},{\bf k'};\omega_n,\omega_n')$, can be calculated using the diagram in Fig. \ref{fig:feynman}(d),
which shows that in the present theory pairing is mediated by charge fluctuations contained 
in $D_{ab}$.
Note that we can also draw a diagram containing two vertices $\Lambda^{pp'}_{ab}$ and two bosonic 
propagators $D_{ab}$, however, this contribution is omitted because it is $O(1/N^2)$.
The analytical expression for the effective interaction is
\begin{equation}
V_{{\rm eff}}({\bf k},{\bf k'};\omega_n,\omega_n')=\Lambda_a D_{ab}({\bf k}-{\bf k'},\omega_n-\omega_n') \Lambda_b \, ,
\label{eq:veff}
\end{equation}
\noindent where $\Lambda_a$ and $\Lambda_b$ are the three-legs vertices from \eq{eq:vertex-lambda} with $p=p'$.

We use a weak coupling approximation to compute
the effective couplings $\lambda_i$
in the different pairing channels or irreducible representations
of the order parameter on the square lattice, $i$ $[i=(d_{x^2-y^2}, d_{xy}, s', p)]$, 
\begin{eqnarray}
\lambda_i=\frac{1}{(2 \pi)^2}
\frac{\int (d {\bf k} /|v_{\bf k}|) \int (d {\bf k'}/|v_{\bf k'}|)
g_i({\bf k'})
V_{{\rm eff}}({\bf k'},{\bf k}) g_i({\bf k})}{
\int (d {\bf k}/|v_{\bf k}|)  g_i({\bf k})^2 }\, ,
\label{eq:lambda-i}
\end{eqnarray}
\noindent
where the functions $g_i({\bf k})$ encode the different pairing
symmetries, $g_{d_{x^2-y^2}}({\bf k})=\cos(k_x)-\cos(k_y)$, 
$g_{d_{xy}}({\bf k})=\cos(k_x)\cos(k_y)$, $g_{s'}({\bf k})=\cos(k_x)+\cos(k_y)$, and $g_{p}({\bf k})=\sin(k_x)$.
$v_{\bf k}$ is the quasiparticle velocity at momentum ${\bf k}$.
The integrations are restricted to the Fermi
surface, i.e., ${\bf k}$ and ${\bf k'}$ run over Fermi surface momenta and ${\rm i}\omega_n={\rm i}\omega'_n=0$.
$\lambda_i$ 
measures the strength of the interaction between electrons at the Fermi surface in a given symmetry channel $i$.
If $\lambda_i > 0$, electrons are repelled hence, superconductivity is only possible
when $\lambda_i < 0$.
The critical temperatures, $T_c$, can then be estimated
using a BCS expression: $T_{ci}= 1.13 \omega_0 \exp(-{1/|\lambda_i|})$,
where $\omega_0$ is a suitable cutoff frequency which 
encodes retardation effects. If $\lambda_i$ is negligible, superconductivity is unexpected, no matter the 
value of $\omega_0$.
Although it is an approximation, the weak coupling scheme 
gives a way to select, in principle, the dominant
pairing channels from all different contributions independently of
their retarded or unretarded nature.
It was introduced in retarded (with glue) cases as the electron-phonon one\cite{Rickaisenbook}, where 
$\lambda$ is the dimensionless coupling strength due to the electron-phonon
interaction. This approach was also 
used for spin-fluctuation interaction in the context of cuprates\cite{monthoux91}.
The fact that we calculate on the Fermi surface in \eq{eq:lambda-i} 
does not invalidate the study of retarded interactions.
Obtaining an accurate value of $T_c$ requires considering retardation effects
in more detail, but that is not our aim.
We study the main tendencies to superconductivity and from where they arise
by computing the coupling strength $\lambda$ of each contribution.

\section{Results and discussions}

We chose $J=0.3$, $T=0$, and $0 \leq V \ll V_c$, where $V_c$ is
the onset of the instability to a checkerboard charge density
wave\cite{foussats04,hoang02}. 
Energy is given in units of $t$.  
There is no tendency to superconductivity, i.e., $\lambda_i > 0$, for any pairing channel except 
for $d_{x^2-y^2}$ for $\delta < 0.5$. Thus, in the following we focus only on the $d_{x^2-y^2}$ channel. For 
simplicity we call $\lambda_{d_{x^2-y^2}}$ ($d_{x^2-y^2}$-wave) as $\lambda$ ($d$-wave) in what follows.

\begin{figure}[!ht]
\begin{center}
\setlength{\unitlength}{1cm}
\includegraphics[width=8cm,angle=0]{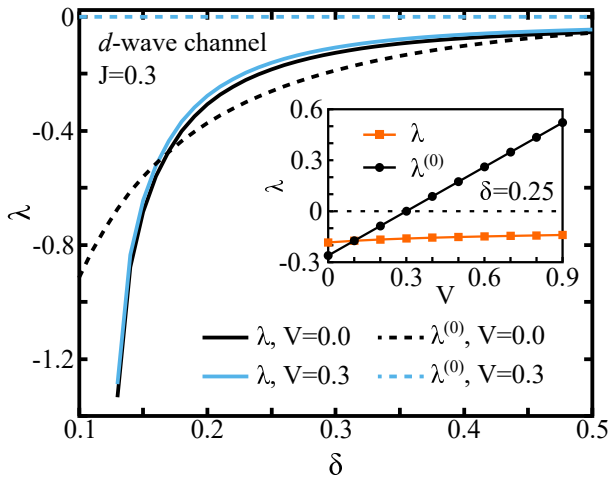}
\end{center}
\caption{(Color online) The superconducting coupling $\lambda$ and $\lambda^{(0)}$  versus doping $\delta$ for 
$V=0$ and $V=0.3$.
Inset: The superconducting coupling $\lambda$ and $\lambda^{(0)}$ versus $V$ 
for $\delta=0.25$. For this $\delta$, $V_c \sim 1.9$.
}
\label{fig:lambda}
\end{figure}

Using the $6 \times 6$ $D_{ab}$ in \eq{eq:veff} we compute $\lambda$ 
as a function  of $\delta$ for $V=0$ and $V=0.3$.
Figure \ref{fig:lambda} shows that, 
although $V(\vq)$ has a repulsive $d$-wave projection, $\lambda$ is almost
unaffected by $V$. In addition, 
$d$-wave superconductivity enhances
with decreasing doping. This result is contrary to other results 
that suggest that superconductivity
is killed by $V$ already for values of 
the order of $J$ (Refs. [\onlinecite{plakida14,zeyher98,zhou04a}]). 

Using the $6 \times 6$ $D_{ab}^{(0)}$ instead of $D_{ab}$ in \eq{eq:veff}
$V_{{\rm eff}}$ is given by\cite{negle}
\begin{align}
V_{{\rm eff}}^{(0)}({\bf k},{\bf k'};\omega_n,\omega_n')=&\left(\frac{J}{2}-V\right)\left[\cos \left(k_x-k'_x\right)+\cos \left(k_y-k'_y\right)\right]+\nonumber\\
			  &\frac{J}{2}\left[\cos \left(k_x-k'_x\right)+\cos \left(k_y-k'_y\right)\right].
\label{eq:veff-D0}
\end{align}

\noindent Note that $V_{{\rm eff}}^{(0)}$ is frequency independent.
The first term in $V_{{\rm eff}}^{(0)}$ containing $J$ and $V$ comes from the $2 \times 2$ sector of the $t$-$J$ model.
The second term originates from the $4 \times 4$ sector which is proportional to $J$. 
We call attention that considering $\tilde{c}$ as usual fermions in \eq{tJV} 
$V_{{\rm eff}}^{(0)}$ can be recovered as a mean-field approximation of the $t$-$J$ model. 

Using $V_{{\rm eff}}^{(0)}$ the corresponding  $\lambda^{(0)}$ can be computed. 
In contrast to $\lambda$, while superconductivity is robust 
for $V=0$, $\lambda^{(0)}$ vanishes for $V=0.3$ (see  Fig. \ref{fig:lambda}). 
These results show that, among other effects discussed later, the renormalization of $D^{(0)}_{ab}$ 
by the $6 \times 6$ boson self-energy $\Pi_{ab}$ [\eq{dyson-A}] screens out the effect of $V$. 
The inset in Fig. \ref{fig:lambda} shows $\lambda$ and $\lambda^{(0)}$ versus $V$ for 
$\delta=0.25$.
These results for $\lambda$ indicate that superconductivity is mostly unaffected by the Coulomb
interaction even for $V \gg J$ when the full dressed $D_{ab}$ bosonic
propagator is considered.
On the other hand, for the case of the bare 
propagator $D^{(0)}_{ab}$ no superconductivity occurs for $V > 0.3$, as can be expected from \eq{eq:veff-D0}.

One difference between $\lambda$ and $\lambda^{(0)}$ for $V=0$ is
that while $\lambda^{(0)}$ smoothly decreases with decreasing doping, $\lambda$ 
tends to large negative values at $\delta \sim 0.13$.
This behavior for $\lambda$ can be explained in the context of the flux phase instability, which occurs 
at a critical doping $\delta_c \sim 0.13$ for present parameters\cite{bejas12}.
See the Appendix for details about the flux phase.
Since $\lambda$ is calculated on the Fermi surface, i.e., $\omega_n=\omega'_n=0$, when approaching 
$\delta_c$ the effective superconducting coupling $\lambda$ tunes the instability and diverges.

\begin{figure}[!ht]
\begin{center}
\setlength{\unitlength}{1cm}
\includegraphics[width=8cm,angle=0]{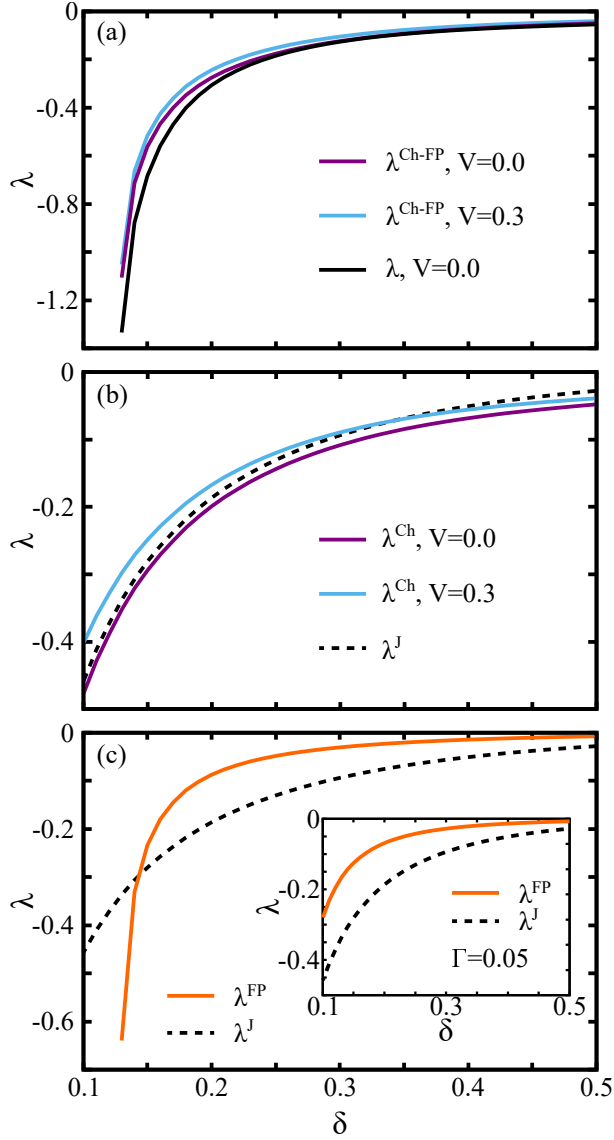}
\end{center}
\caption{(Color online)
(a) $\lambda^{{\rm Ch-FP}}$ versus doping for $V=0$ and $V=0.3$. $\lambda$ from Fig.~\ref{fig:lambda} is included
for comparison. (b) $\lambda^{{\rm Ch}}$ for $V=0$ and $V=0.3$, and $\lambda^{\rm J}$ versus doping.
(c) $\lambda^{{\rm FP}}$ and $\lambda^{\rm J}$ versus $\delta$. Inset: $\lambda^{{\rm FP}}$ and $\lambda^{\rm J}$ versus $\delta$ for $\Gamma=0.05$.  
 }
\label{fig:paneles}
\end{figure}

It was shown that $\lambda$, which includes the bosonic
self-energy $\Pi_{ab}$, is robust against $V$, but such robustness is not present in the case of $\lambda^{(0)}$. 
Next, we discuss which are the relevant components of $\Pi_{ab}$ that lead to the different behavior 
between $\lambda$ and $\lambda^{(0)}$. Since the flux phase belong to the $5$-$6$ sector of $D_{ab}$ (see the Appendix), 
we calculate $\lambda$ including only the $2 \times 2$ sector $\Pi_{11}$, $\Pi_{12}$, $\Pi_{22}$, and 
the flux sector $\Pi_{55}$, $\Pi_{56}$ and $\Pi_{66}$ in 
the Dyson equation [\eq{dyson-A}], i.e.,
leaving the other components of $\Pi_{ab}$ as zero.
We call this $\lambda^{{\rm Ch-FP}}$.
Figure \ref{fig:paneles}(a) shows $\lambda^{{\rm Ch-FP}}$ for $V=0$ and $V=0.3$. 
For completeness, in the figure we included the results of $\lambda$ for the
full $6 \times 6$ case of Fig.~\ref{fig:lambda}.
These results show that $\Pi_{11}$, $\Pi_{12}$, $\Pi_{22}$, $\Pi_{55}$, $\Pi_{56}$, and $\Pi_{66}$ 
are the most important components of the bosonic self-energy $\Pi_{ab}$ since they 
capture the same $\lambda$ behavior as using the full $6 \times 6$ $\Pi_{ab}$.
It is important to note that the inclusion of $\Pi_{ab}$ in the Dyson equation introduces a frequency dependence in
the dressed bosonic propagator $D_{ab}$, i.e., the effective interactions are retarded
in contrast to the unretarded interactions from the undressed $D^{(0)}_{ab}$.
This point is important for later discussions.

Next we analyze the influence of the $2 \times 2$ on-site-charge and FP sectors 
separately.

Considering only $\Pi_{11}$, $\Pi_{12}$, and $\Pi_{22}$ in the Dyson equation 
the effective paring interactions can be written as

\begin{align}
V_{eff}^{({\rm Ch})}({\bf k},{\bf k'};\omega_n,\omega_n')=&
\frac{-2\Lambda_1(\delta-\Pi_{12})+\Lambda_1^2\Pi_{22}-\{\frac{\delta^2}{2}(2V-J)F_{\vk,\vk'}-
\Pi_{11}\}}{(\delta-\Pi_{12})^2+\Pi_{22}\{\frac{\delta^2}{2}(2V-J)F_{\vk,\vk'}-\Pi_{11}\}}+\nonumber\\
			  &\frac{J}{2}\left[\cos \left(k_x-k'_x\right)+\cos \left(k_y-k'_y\right)\right],
\label{eq:veff-only12}
\end{align}
\noindent where $F_{\vk,\vk'} = \cos \left(k_x-k'_x\right)+\cos \left(k_y-k'_y\right)$.
Using \eq{eq:veff-only12}, we compute $\lambda^{{\rm Ch}}$.
Figure \ref{fig:paneles}(b) shows results for $\lambda^{{\rm Ch}}$ versus $\delta$  
for $V=0$ and $V=0.3$.
It can be seen that $\lambda^{{\rm Ch}}$ is almost unaffected by $V$ showing that the Coulomb 
repulsion is indeed screened by the $\Pi_{ab}$ components that belong to the $2 \times 2$ 
on-site charge sector. 
The second term on the right hand side of \eq{eq:veff-only12} is the same as in \eq{eq:veff-D0}.
We call $\lambda^{J}$ the contribution from this term and its behavior is shown in
Fig. \ref{fig:paneles}(b).
The fact that the three curves are nearly coincident give us the clue that
the components of the $2\times2$ on-site charge sector of $\Pi_{ab}$ screen
the first term of \eq{eq:veff-D0} and consequently, only the effective 
$(J/2) \, \left[\cos \left(k_x-k'_x\right)+\cos \left(k_y-k'_y\right)\right]$
interaction from the $4\times4$ sector survives. 

When the $t$-$J$ model is treated at mean-field
level, superconductivity is expected to be triggered by the exchange term $J({\bf k}-{\bf k'})$.
However, superconductivity is killed by a small nearest-neighbors Coulomb interaction. 
When the non-double-occupancy constraint is treated properly, the effect of the Coulomb interaction is screened.
Thus, present results show a clear difference between a treatment of superconductivity at the mean-field 
level and a treatment in strong coupling. 
We think that our results support the early point of view\cite{anderson07} that superconductivity 
in cuprates has a contribution from the 
unretarded $J({\bf k}-{\bf k'})$ term, but we claim that for such a pairing to be realized 
the non-double-occupancy constraint should be treated beyond mean-field. 

The screening effect from the $2 \times 2$ sector can be understood as follows. 
The second contribution of the first term in \eq{eq:veff-only12} is mainly $s$-wave and gives a negligible
contribution in the $d$-wave channel, i.e., this term is not relevant for our analysis. 
The third term has the form 
of the screening of the Coulomb interaction from the usual RPA, because $\Pi_{22}$ is just 
a simple bubble. It is important to remember that $\Pi_{22}$ arises here from  fluctuations 
of the Lagrange multiplier introduced to impose the constraint.
Then, this contribution screens the $J$ and $V$ terms (first term of $V_{{\rm eff}}^{(0)}$) 
from the $2 \times 2$ sector, while the $J$-term from the $4 \times 4$ sector (second term in $V_{{\rm eff}}^{(0)})$ 
remains. In addition, the first contribution $-2\Lambda_1(\delta-\Pi_{12})$, which is independent of $J$ and $V$ 
has a small repulsive $d$-wave projection.
Then, if $V=J=0$, i.e., only the $2\times2$ sector is present, 
superconductivity is not expected to be mediated by charge fluctuations.

It is important to mention that the doping dependence of  $\lambda^{{\rm Ch}}$ and $\lambda^{\rm J}$ does not show 
the steep behavior near $\delta_c$ seen in Fig. \ref{fig:lambda} 
for $\lambda$.
This is due to the fact that we did not include $\Pi_{55}$, $\Pi_{56}$ and $\Pi_{66}$ 
from the FP sector
(see the Appendix). 
To understand the influence of only these components on $\lambda$ we
take the dressed bosonic propagator $D_{ab}$ and compute $V_{{\rm eff}}$ 
by projecting $D_{ab}$ onto the FP eigenvector $(0,0,0,0,1/\sqrt{2},-1/\sqrt{2})$ [Ref.\onlinecite{bejas12}]. 
We obtain
\begin{equation}
V_{{\rm eff}}^{({\rm FP})}\left({\bf k},{\bf k'};\omega_n,\omega_n'\right)=
-(\Lambda_5-\Lambda_6)^2{\rm Re}\chi_{{\rm FP}}({\bf k}-{\bf k'},{\rm i}\omega_n-{\rm i}\omega'_n)\, ,
\label{eq:veff-fp}
\end{equation}
\noindent where $\Lambda_5$ and $\Lambda_6$ are the fifth and sixth component of the vertices in \eq{eq:vertex-lambda}, and
\begin{eqnarray}
\chi_{{\rm FP}}(\vq,{\rm i}\nu_n)=
 [(8/J) \Delta^2-\Pi_{{\rm FP}}(\vq,{\rm i}\nu_n)]^{-1}\,,
\label{chi-flux}
\end{eqnarray}
\noindent which is the flux phase susceptibility\cite{bejas12} and 
$\Pi_{{\rm FP}}(\vq,{\rm i}\nu_n)$ the electronic polarizability 
given by 
\begin{eqnarray}
\Pi_{{\rm FP}}(\vq, {\rm i}\nu_n) = - \frac{1}{N_{s}}\;
\sum_{\vk}\; \gamma_{{\rm FP}}^2(\vq,\vk) \frac{n_{F}(\epsilon_{\vk + \vq}) 
- n_{F}(\epsilon_{\vk})} 
{\epsilon_{\vk + \vq} - \epsilon_{\vk}-{\rm i} \nu_n}\,,
\label{PiCDW}
\end{eqnarray}
\noindent with the form factor 
$\gamma_{{\rm FP}}(\vq,\vk)=2 \Delta [\sin(k_x+q_x/2)-\sin(k_y+q_y/2)]$. 
For $\vq=(\pi,\,\pi)$ the form factor $\gamma_{{\rm FP}}(\vq,\vk)$ transforms as $[\cos(k_x)-\cos(k_y)]$, i.e., 
the flux instability has $d$-wave symmetry. $\chi_{{\rm FP}}(\vq,{\rm i}\nu_n)$ plays the role of a 
bosonic glue, as phonons in usual superconductors.

This projection isolates the FP sector and allows us to check its effect on $\lambda$.
$\lambda^{{\rm FP}}$ versus $\delta$, where $\lambda^{{\rm FP}}$ is calculated using $V_{{\rm eff}}^{({\rm FP})}$, 
is shown in Fig. \ref{fig:paneles}(c). While at large doping $\lambda^{{\rm FP}}$ goes to zero, the curve
shows the steep behavior approaching $\delta_c$. 
In Fig. \ref{fig:paneles}(c), we also plot $\lambda^{\rm J}$ versus $\delta$.
Comparing $\lambda^{{\rm FP}}$ with $\lambda^{\rm J}$ we conclude that the flux phase enhances superconductivity only near
the quantum critical point at $\delta_c$ associated with the flux instability. 
This tendency to enhance superconductivity can be seen as triggered by 
quantum critical fluctuations\cite{abanov19}.

Figure \ref{fig:paneles} shows that the total coupling strength 
$\lambda$ can be computed in a good approximation as the sum of $\lambda^{{\rm FP}}$ and $\lambda^{\rm J}$, i.e., as 
coming from an effective pairing interaction $V_{{\rm eff}} \sim V_{{\rm eff}}^{({\rm FP})} + J(\vk-\vk')$.
While $V_{{\rm eff}}^{({\rm FP})}$ is retarded, $J(\vk-\vk')$ is unretarded. 
It is well known that when introducing a 
finite broadening $\Gamma$
in the analytical continuation ${\rm i}\nu_n = \nu + {\rm i} \Gamma$ in \eq{PiCDW},
the flux phase is pushed 
to lower dopings\cite{yamase19}. In the inset of Fig. \ref{fig:paneles}(c) we show results 
for $\lambda^{{\rm FP}}$ and $\lambda^{\rm J}$ for $\Gamma=0.05$. For this $\Gamma$ 
the flux-phase does not set down at a finite doping, and $J(\vk-\vk')$ is a good approximation 
for computing the total coupling strength for all dopings.  

The authors of Ref.[\onlinecite{maier08}]
showed that the pairing strength is composed by a retarded spin fluctuation contribution 
and an unretarded term $J(\vk-\vk')$, and that the retarded pairing dominates. 
In agreement with this work 
we also found an unretarded $J(\vk-\vk')$ contribution. 
As discussed in our paper the large-$N$ 
approximation weakens spin fluctuations over charge 
fluctuations, then we cannot rule out the presence of a retarded spin-fluctuation pairing. 
In Ref. [\onlinecite{maier08}] the Coulomb potential $V(\vq)$ was not included, which can kill the superconductivity 
from the spin fluctuation term. However, we showed that the
constraint in the $t$-$J$ model, when included, screens $V(\vq)$. Then, we think that our paper and
that of Ref. [\onlinecite{maier08}] are complementary. If pairing in cuprates is mainly
retarded or mainly unretarded is an open discussion.
Although one can expect a retarded pairing as in conventional superconductors,
some experiments suggest that pairing may certainly be
unretarded\cite{lorenzana13,park13}.

\section{Conclusions}

Using a large-$N$ approach on the microscopic $t$-$J$-$V$ model we studied $d$-wave superconductivity 
and the role 
of a nearest-neighbors Coulomb repulsion on it. In this approach, pairing is mediated by a 
bosonic propagator which contains 
on-site charge 
and bond-charge fluctuations, both treated at the same footing in present formalism. 
When the bare bosonic propagator is 
considered, superconductivity arises from  the unretarded exchange term 
$J({\bf k}-{\bf k'})$. 
However,  
the presence of the nearest-neighbors Coulomb repulsion 
$V({\bf q})$ is detrimental to superconductivity and cancels pairing for values 
of $V \sim J$, suggesting a fragile $d$-wave superconductivity. 
The situation changes drastically when the bosonic propagator is dressed by interactions.
In this case, 
superconductivity becomes almost 
unaffected by $V$ and remains robust even for $V \gg J$. 
The inclusion of the non-double-occupancy
constraint and its fluctuations screens the effect of $V({\bf q})$, while a pairing contribution from the
$J$-term remains.   
In other words, the scenario for a possible unretarded (without glue)  
pairing contribution triggered by $J$ emerges in strong coupling, i.e., only if the local
constraint is considered properly.  

Our results may be useful for the comparison with similar calculations in the Hubbard and $t$-$J$ models.
A robust $d$-wave superconductivity against a nearest-neighbors Coulomb 
repulsion $V({\bf q})$ requires the non-double occupancy to be considered, and at this level 
an unretarded pairing contribution is obtained. 
In the large-$U$ limit, the Hubbard model is mapped to the $t$-$J$ model. Then it 
would be interesting to check the role of $V({\bf q})$ on superconductivity and, in addition, 
to disentangle retarded and unretarded interactions from the obtained pairing in the Hubbard model.

\acknowledgments
The authors thank P. Bonetti, W. Metzner, D. Vilardi, and H. Yamase for fruitful discussions. 
A.G. thanks the Max-Planck-Institute for Solid State Research in Stuttgart for hospitality and 
financial support. 

\appendix
\newpage

\section{Some characteristics and discussions on the flux phase}

In this Appendix we briefly discuss
the main characteristics of the flux phase (FP) and its possible connection with the physics of the pseudogap.
As discussed in Ref. [\onlinecite{bejas12}], for present parameters ($J=0.3$ and $T=0$) 
the flux phase\cite{affleck88a,affleck89,morse91,cappelluti99,foussats04} 
occurs at $\delta=\delta_c \sim 0.13$, 
with a modulation vector ${\bf Q}$ 
close to $(\pi,\pi)$, i.e., the FP breaks the translational symmetry. In present large-$N$ approximation
the FP occurs when one eigenvalue of $D^{-1}_{ab}$ is zero, and since the associated 
eigenvector is of the form $(0,0,0,0,1/\sqrt{2},-1/\sqrt{2})$, the flux instability 
is located in the sector $5$-$6$  of the $6 \times 6$ matrix $D_{ab}$ (Ref.[\onlinecite{bejas12}]). 
For $\delta <\delta_c$ the imaginary 
components $A^x$ and $A^y$ of the bond field become finite. 
The commensurate FP is characterized by the 
modulation vector $\vq=(\pi,\,\pi)$ and describes staggered circulating currents.
In the FP state a $d$-wave gap, similar to the pseudogap in cuprates, 
opens, and Fermi pockets with 
low intensity in the outer part are developed\cite{chakravarty03} instead of a large Fermi surface.
The FP is equivalent to the $d$CDW
which was proposed phenomenologically for describing the pseudogap\cite{chakravarty01}.

The FP is a bond-charge instability. As discussed
in Ref. [\onlinecite{bejas12}], besides the FP there are several kinds of bond-charge fluctuations, 
and in principle all of them can lead to an instability depending on the model parameters.
However, in the context of the present large-$N$ method, for hole-doped cuprates it was found that
the flux instability is robust in a realistic-parameters regime. In contrast, 
for electron-doped cuprates the leading bond-charge instability can occurs for the real components $r^x$ 
and $r^y$ of the fluctuations of the bond field \cite{bejas14}.
 
Although the flux phase or $d$CDW is a candidate for 
describing the pseudogap,
its existence 
in the $t$-$J$ and Hubbard models is controversial. 
While some reports show the presence of the flux instability or its 
fluctuations\cite{leung00,dong20}, others do not\cite{macridin04}. 
The FP is also controversial from the experimental point of view.
While the authors of Refs. [\onlinecite{chakravarty01,honerkamp04,lee06}] show that a
series of experiments in the 
pseudogap phase can be described in the context of the FP,
angle-resolved photoemission spectroscopy
(ARPES) experiments do not show pockets but Fermi arcs\cite{norman98,damascelli03} which are 
considered as an indication that 
translational symmetry is not broken  in the pseudogap. 
In Refs. [\onlinecite{greco09,greco11,greco14}]  
the interaction between the flux-phase
fluctuations and carriers in the proximity to the flux-phase instability leads to a reasonable description
of the Fermi arcs and Raman scattering without the necessity of the translational-symmetry breaking. 
Recently\cite{gourgout20}, it was  proposed that the FP is a good candidate for describing 
the pseudogap.

The connection between the FP and the antiferromagnetism and its fluctuations,
which lead to $d$-wave superconductivity\cite{senechal13,vilardi19}, is an interesting point.
The FP occurs at much larger doping ($\delta=0.13$ in the present calculation) than 
the onset of antiferromagnetism. Then, at the onset of the FP both phases may interact weakly while, 
with decreasing doping approaching the antiferromagnetic-insulating phase, antiferromagnetism and its 
fluctuations may
lead against the FP. On the other hand, the FP 
develops staggered magnetic moments much weaker than those in the 
antiferromagnetic phase\cite{hsu91} which, in principle, indicates that the FP and antiferromagnetism are distinct 
phases.
In spite of that, 
it was claimed that antiferromagnetism can be also  understood in the framework of the flux phase\cite{ho01}.

\newpage

\bibliography{main1} 

\begin{thebibliography}{51}
\bibitem[{neg()}]{negle}
\bibinfo{note}{A term proportional to $D^{(0)}_{12}$ was neglected because does
  not contribute to the $d$-wave channel.}
\end{thebibliography}

\end{document}